\documentclass[conference,10pt]{IEEEtran}
\IEEEoverridecommandlockouts
\usepackage{cite}
\usepackage{amsmath,amssymb,amsfonts}
\usepackage{algorithmic}
\usepackage{graphicx}
\usepackage{textcomp}
\usepackage{xcolor}
\usepackage{url}
\usepackage{multirow}
\usepackage{multicol}
\usepackage[capitalise,nameinlink,noabbrev]{cleveref}
\usepackage[vlined, boxed, ruled, linesnumbered]{algorithm2e}
\usepackage{booktabs} 
\usepackage{etoolbox}
\usepackage{listings}
\usepackage{subfigure}
\usepackage{dcolumn}
\newcolumntype{d}[1]{D{.}{.}{#1}}
\newcommand\mc[1]{\multicolumn{1}{c}{#1}} 
\newcommand\mcl[1]{\multicolumn{1}{l}{#1}} 
\usepackage[varqu]{zi4}

\lstdefinestyle{x86asm}{
    float=tp,
    floatplacement=tbp,
    frame=single, 
    xleftmargin=\parindent,
    language=[x86masm]Assembler,
    captionpos=b,
    basicstyle=\footnotesize\ttfamily,
    commentstyle=\itshape\color{green!60!black},
    keywordstyle=\color{blue!80!black},
    identifierstyle=\color{red!80!black},
    tabsize=4,
    numbers=left,
    numbersep=6pt,
    stepnumber=1,
    numberstyle=\tiny\color{gray}, 
    columns=fullflexible,
    morekeywords={vxorps,vbroadcastss,vfmadd231ps,vfmadd231ss,vmovups,vmovss},
}

\usepackage{color}
\newcommand{\thomas}[1]{}
\renewcommand{\thomas}[1]{}
\newcommand{\qiang}[1]{}
\renewcommand{\qiang}[1]{}
\newcommand{\howie}[1]{}
\renewcommand{\howie}[1]{{\color{cyan} From Howie: {#1}}}

\def\BibTeX{{\rm B\kern-.05em{\sc i\kern-.025em b}\kern-.08em
    T\kern-.1667em\lower.7ex\hbox{E}\kern-.125emX}}

\newcommand{\name}{{\textsc{JitSpMM}}\xspace}

\begin{document}

\title{{\name}: Just-in-Time Instruction Generation for Accelerated Sparse Matrix-Matrix Multiplication}


\author{\IEEEauthorblockN{Qiang Fu\textsuperscript{\textasteriskcentered}}
\IEEEauthorblockA{
\textit{Advanced Micro Devices Inc.}\\
Austin, TX USA \\
\url{charlifu@amd.com}}
\and
\IEEEauthorblockN{Thomas B. Rolinger\textsuperscript{\textsection}}
\IEEEauthorblockA{
\textit{NVIDIA}\\
Austin, TX USA \\
\url{trolinger@nvidia.com}}
\and
\IEEEauthorblockN{H. Howie Huang}
\IEEEauthorblockA{
\textit{George Washington University}\\
Washington, DC USA \\
\url{howie@gwu.edu}}
}

\maketitle

\begingroup\renewcommand\thefootnote{\textasteriskcentered}
\footnotetext{Work completed at George Washington University}
\endgroup

\begingroup\renewcommand\thefootnote{\textsection}
\footnotetext{Work completed at the Laboratory for Physical Sciences.}
\endgroup

\begin{abstract}
Achieving high performance for Sparse Matrix-Matrix Multiplication (SpMM) has received increasing research attention, especially on multi-core CPUs, due to the large input data size in applications such as graph neural networks (GNNs).
Most existing solutions for SpMM computation follow the ahead-of-time (AOT) compilation approach, which compiles a program entirely before it is executed.
AOT compilation for SpMM faces three key limitations: unnecessary memory access, additional branch overhead, and redundant instructions.
These limitations stem from the fact that crucial information pertaining to SpMM is not known until runtime.
In this paper, we propose {\name}, a just-in-time (JIT) assembly code generation framework to accelerated SpMM computation on multi-core CPUs with SIMD extensions. 
First, {\name} integrates the JIT assembly code generation technique into three widely-used workload division methods for SpMM to achieve balanced workload distribution among CPU threads.
Next, with the availability of runtime information, {\name} employs a novel technique, \emph{coarse-grain column merging}, to maximize instruction-level parallelism by unrolling the performance-critical loop.
Furthermore, {\name} intelligently allocates registers to cache frequently accessed data to minimizing memory accesses, and employs selected SIMD instructions to enhance arithmetic throughput.
We conduct a performance evaluation of {\name} and compare it two AOT baselines.  
The first involves existing SpMM implementations compiled using the Intel \emph{icc} compiler with auto-vectorization. 
The second utilizes the highly-optimized SpMM routine provided by Intel MKL.
Our results show that {\name} provides an average improvement of 3.8\texttimes~and 1.4\texttimes, respectively.
\end{abstract}

\section{Introduction}
\begin{figure}[t]
    \centering
    \includegraphics[width=0.87\linewidth]{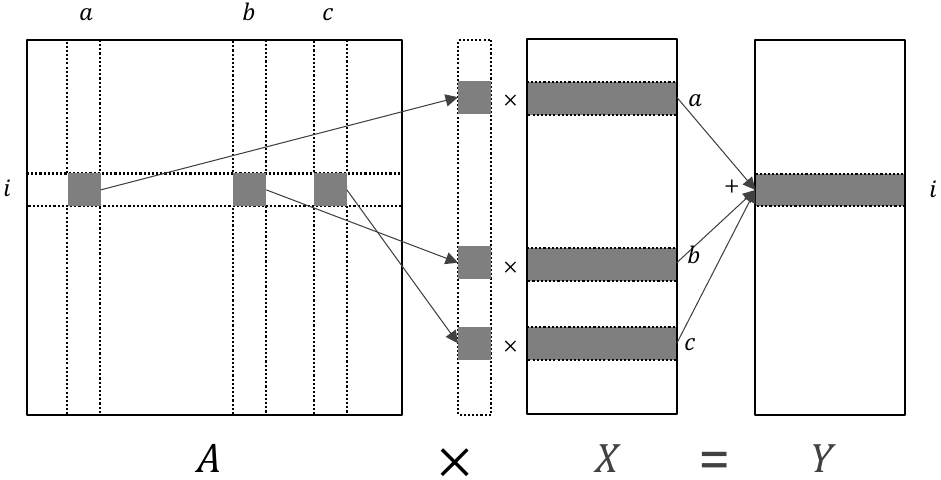}
    \caption{The process of calculating a single row $i$ for the SpMM operation. In this case, the $i$-th row of the sparse matrix $A$ has three non-zero elements at columns $a$, $b$, and $c$. To get the result, these three values are multiplied with each element in the corresponding rows of matrix $X$ (at indices $a$, $b$, and $c$) and summed up into one as the result, which is the $i$-th row of $Y$.}
    \label{fig:spmm}
\end{figure}

The growing attention on \emph{sparse computation} in the deep learning community makes it a natural focus for performance optimization\cite{bik2022compiler,hegde2019extensor,ye2023sparsetir,chou2022compilation}, given that sparse operations can exploit the naturally occurring sparsity in data to reduce storage requirements and computational time\cite{bik2022compiler}.
\emph{\textbf{SpMM (Sparse Matrix-Matrix Multiplication)}} is a widely used sparse operation, which involves the multiplication of a sparse matrix $A$ by a dense matrix $X$, resulting in a dense matrix $Y$. 
The process of computing a single row for the SpMM operation is depicted in Figure \ref{fig:spmm}.
For the calculation of the $i$-th row of $Y$, the non-zero values in the $i$-th row of the sparse matrix $A$ are multiplied with the corresponding rows in the dense matrix $X$ and then aggregated to form the result.
SpMM is commonly used in algorithms such as PageRank\cite{langville2006google}, matrix factorization\cite{koren2009matrix}, and graph clustering\cite{schaeffer2007graph}.
Additionally, SpMM is a key component of the graph convolution, which is a fundamental operation in Graph Neural Networks (GNNs) \cite{abadal2021computing,wang2019deep,fu2022tlpgnn,fu2021automatic,yin2023dgi}.
By efficiently computing the graph convolution using SpMM, GNNs can scale to handle large graphs and achieve state-of-the-art performance in various tasks such as node classification\cite{rong2019dropedge}, link prediction\cite{zhang2018link}, and graph classification\cite{zhou2020graph,dwivedi2023benchmarking,fu2018combating}.

Existing solutions for SpMM computation, including Intel MKL\cite{wang2014intel}, primarily follow an \textbf{\emph{ahead-of-time (AOT) compilation}} approach\cite{yang2018design}. 
In this approach, the SpMM operation is implemented using low-level coding (C/C++ or assembly) and compiled into an executable binary that can handle inputs of varying sizes\cite{merrill2016merge, wang2014intel}.
\thomas{Do you mean high-level languages? I think that is the usual workflow for AOT, taking C/C++ and producing machine code.}\qiang{I think we can go with low-level coding (C/C++ or assembly).}
Most optimization efforts are directed at the source code level, with techniques such as merge-based workload division\cite{merrill2016merge} proposed to achieve better workload balance for SpMM\cite{yang2018design}. 
However, AOT approaches face three major limitations: 1) \emph{Unnecessary memory access} because of the register allocation which is not optimized for SpMM computation, 2) \emph{additional branch overhead} due to the lack of runtime information utilization, and 3) \emph{redundant instructions} executed for unnecessary memory access and branch control.

To overcome the above limitations, we propose the adoption of \textbf{\emph{just-in-time (JIT) assembly code generation}} which involves directly issuing assembly codes for SpMM computation at runtime.
JIT assembly code generation enables us to optimize the SpMM operation at the assembly level, allowing for a more tailored register allocation that minimizes unnecessary memory accesses. 
By leveraging runtime information, the JIT approach efficiently circumvents the branch overhead inherent in AOT solutions. 
As a result, JIT assembly code generation reduces the amount of redundant instructions during SpMM by eliminating unneeded memory accesses and branch control.
Our experiment on single-thread scalar SpMM shows that JIT assembly code generation can reduce the memory loads by 2.4 - 2.7\texttimes, branch misses by 1.2 - 4.1\texttimes, and instruction execution by 3.4 - 4.4\texttimes, when compared to three AOT C/C++ compilers.

\begin{table}[t]
    \centering
    \caption{List of notations.}
    \begin{tabular}{c|l}
    \toprule
        \textbf{Symbol} & \textbf{Description}  \cr \midrule
        $A$ & Sparse matrix of dimensions: $m \times n$ \cr
        $X$ & Input dense matrix of dimensions: $n \times d$ \cr
        $Y$ & Output dense matrix of dimensions: $m \times d$ \cr
        $m$ & The number of rows of matrix $A$ \cr
        $n$ & The number of columns of matrix $A$ \cr
        $d$ & The number of columns of input dense matrix $X$ \cr
        $M[i]$ & The $i$-th row of a matrix $M$ \cr
        $M[i][j]$ & The element at the $i$-th row and $j$-th column of a matrix $M$ \cr
        $v[i\colon j]$ & The elements in the slice $[i,j)$ of a vector $v$  \cr
    \bottomrule
    \end{tabular}
    \label{tab:notation}
\end{table}

In this paper, we introduce {\name}, a JIT assembly code generation framework designed to accelerate SpMM computation on multi-core CPUs with SIMD instruction extensions. 
First, {\name} integrates the JIT assembly code generation technique into three widely-used workload division methods for SpMM\cite{odeh2012merge} to achieve balanced workload distribution among CPU threads.
Second, with the availability of runtime information, {\name} employs a novel technique, \emph{coarse-grain column merging}, to maximize instruction-level parallelism by unrolling the performance-critical loop.
Finally, {\name} intelligently allocates registers to cache frequently accessed data, minimizing accesses to memory, and employs selected SIMD instructions to enhance arithmetic throughput.

We implement {\name} and assess its performance through a comparative analysis with two state-of-art AOT methods. 
The first baseline is the C++ SpMM implementation derived from the work by Merrill and Garland\cite{merrill2016merge}, which is compiled using Intel \texttt{icc} with auto-vectorization. 
The second baseline consists of the highly optimized SpMM routine from Intel MKL. 
Across an extensive array of matrix datasets, our {\name} consistently outperforms both baselines with an average speedup of 3.8\texttimes~and 1.4\texttimes, respectively. 
Furthermore, through comprehensive profiling results, we illustrate that {\name} effectively reduces unnecessary memory accesses, branch operations, and overall instructions.
This work makes the following primary contributions:
\begin{itemize}
    \item We identify the limitations of AOT approaches that impede improved performance for SpMM computation.
    \item We propose {\name} to address these limitations by adopting JIT assembly code generation for SpMM on multi-core CPU with SIMD extension.
    \item We implement {\name} and demonstrate the effectiveness through a comprehensive performance benchmark and profiling analysis.
\end{itemize}
\thomas{While it isn't a hard requirement, we may want to consider transforming the above paragraph into the typical contributions list. When I read papers, I tend to hone in on this list in the intro to get a feel for what the paper will be presenting.}

The remainder of this paper is structured as follows:
\cref{sec:bkg} provides an introduction to SpMM operations and modern multi-core CPUs.
We present the motivation to adopt JIT assembly code generation in \cref{sec:moti}.
\cref{sec:jitcodegen} elaborates on the details of our {\name} framework and \cref{sec:exp} describes the experiments conducted to evaluate its performance.
We present a brief survey of related works in \cref{sec:rel}.
Finally, \cref{sec:con} offers concluding remarks.


\section{Background}
\label{sec:bkg}

\subsection{Sparse Matrix-Matrix Multiplication (SpMM)}

\begin{figure}[t]
    \centering
    \includegraphics[width=\linewidth]{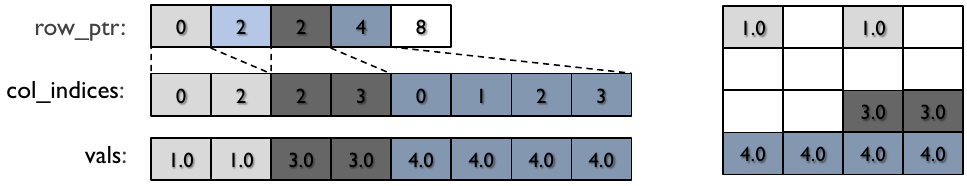}
    \caption{Example of CSR format (left) and the matrix represented (right).}
    \label{fig:csr}
\end{figure}

In general, SpMM is formulated as $Y = AX$, where $A$ is a sparse matrix with dimension $m\times n$, $X$ is a dense matrix with dimension $n\times d$ and, $Y$ is the dense result matrix wih dimension $m \times d$.
There are a variety of formats to represent sparse matrices, but we focus on Compress Sparse Row (CSR) to store the sparse matrix $A$ because it is widely used in vendor libraries (e.g., Intel MKL \cite{wang2014intel}), data science toolkits (e.g., SciPy \cite{virtanen2020scipy}), and GNN frameworks (e.g., DGL \cite{wang2019deep}, PyG \cite{fey2019fast}).
The CSR format stores a sparse matrix in three 1-D arrays: \texttt{row\_ptr}, \texttt{col\_indices}, and \texttt{vals}, as shown in \cref{fig:csr}.
To optimize storage space, CSR efficiently packs the column indices and values of all non-zero elements into \texttt{col\_indices} and \texttt{vals} arrays, respectively, in row-order.
The \texttt{row\_ptr} array stores the offsets of each row's first non-zero element into the other two arrays.
Furthermore, it is noteworthy that in real-world applications like GNNs, the shape of the input dense matrix $X$ often leans towards being "tall and skinny," indicating that $n$ is significantly greater than $d$\cite{selvitopi2021distributed}.

\begin{algorithm}[t]
\footnotesize
\caption{A Sequential Implementation of SpMM based on CSR}\label{alg:seqcsr}
\SetKwInOut{Input}{Input}
\SetKwInOut{Out}{Output}

\Input{Spare matrix $A$ of size $m\times n$, i.e., $A.row\_ptr$, $A.col\_indices$, $A.vals$, and dense matrix $X$ of size $n \times d$.}
\Out{Dense matrix $Y=AX$.}
    \SetInd{0.1em}{0.6em}
    \For{$i = 0\ \textbf{to}\ m$} {
        \For{$j = 0\ \textbf{to}\ d$}{ 
            $ret = 0$ \\
            \For{$idx = A.row\_ptr[i]\ \textbf{to}\ A.row\_ptr[i+1]$} {
                $k = A.col\_indices[idx]$ \\
                $ret \mathrel{+}= A.vals[idx] * X[k][j]$
            }
            $Y[i][j] = ret$
        }
    }
    \Return{$Y$}
\end{algorithm}

Algorithm \ref{alg:seqcsr} illustrates the computation of SpMM.
Each element of the output $Y[i][j]$ is the dot-product between the sparse row $i$ of $A$ and the dense column $j$ of $X$.
The algorithm accesses the $A.row\_ptr$ array to determine the starting position of row $i$ and it iterates over a segment of the $A.col\_indices$ and $A.vals$ arrays to process all the non-zero elements in sparse row $i$.
For each non-zero element, the algorithm utilizes the corresponding column index from $A.col\_indices$ (denoted as $k$ on line 5) to locate the corresponding row in the dense matrix $X$. 
At this point, the algorithm retrieves the value of $X[k][j]$, multiplies it by the non-zero element value from $A.vals$, and adds the result to the accumulated value $ret$. 
This process continues until all non-zero elements in sparse row $i$ have been processed, resulting in the final value of $Y[i][j]$.
\thomas{Maybe one thing to add somewhere in this subsection is why SpMM is challenging for performance, which would be something about the irregular memory accesses.}

\subsection{Modern Multi-Core CPUs and SIMD Instruction Extension}

\begin{figure}[!t]
    \centering
    \includegraphics[width=0.85\linewidth]{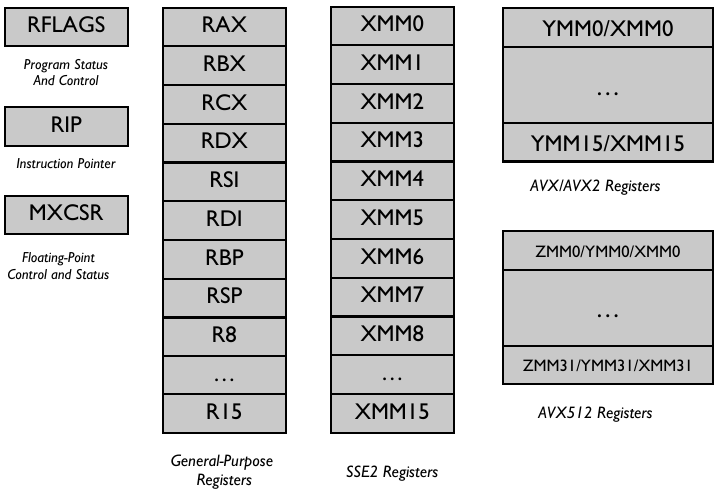}
    \caption{x86-64 processor internal architecture.}
    \label{fig:cpuarch}
\end{figure}

Graphics processing units (GPUs) and domain-specific accelerators have been used to perform sparse operations, such as SpMM, by leveraging their immense computational capabilities and high memory bandwidth \cite{zhu2019sparse, yang2018design, huang2020ge, ortega2014fastspmm}.
However, there is still significant value in exploring techniques to optimize SpMM on general-purpose CPUs.
This is because CPUs continue to serve as the primary computing resource in many systems, offering substantial memory capacity \cite{gong2022graphite,liu2019optimizing}.
For instance, when performing SpMM on the \emph{com-Friendster} dataset\footnote{\url{https://sparse.tamu.edu/SNAP/com-Friendster}} with an input dense matrix with 32 columns, Intel MKL\cite{wang2014intel} consumes 87.2 GB of memory, surpassing the memory capacity of many GPUs available today (e.g., Nvidia's A100 with 80 GB of memory).
Therefore, the focus of this work is to enhance the performance of SpMM for large matrices that often exceed the memory capacity of accelerators.

Modern CPUs leverage thread-level parallelism through multi-core architectures to enhance overall performance, addressing the limitations in scaling up single-core processors due to transistor budget constraints\cite{olukotun1996case}.
Regarding the interface between software and hardware, many modern CPUs utilize the x86 instruction set architecture (ISA).
From the perspective of an executing program, a 64-bit x86 processor's internal architecture can be conceptually divided into several distinct units, including instruction pointer (RIP register), general-purpose registers, status and control flags (RFLAGS register), floating-point registers, control and status (MXCSR), and SIMD extension, as illustrated in Figure \ref{fig:cpuarch}. 
Of particular relevance to our work, SIMD (Single Instruction, Multiple Data) instructions enhance performance by processing multiple data elements simultaneously using vector registers.
The SSE2 extension introduces 16 128-bit registers (\emph{XMM0--15}), while the AVX/AVX2 extension extends the register size to 256 bits (\emph{YMM0--15}).
The most recent extension, AVX512, provides 32 larger 512-bit registers (\emph{ZMM0--31}), allowing for parallel processing of up to 16 different 32-bit floating-point numbers.
For example, the instruction \texttt{vmulps} \texttt{zmm2}, \texttt{zmm1}, \texttt{zmm0} performs a multiplication operation on each pair of single-precision values within \emph{ZMM0} and \emph{ZMM1}, subsequently storing the results in \emph{ZMM2}.



\section{Motivations}
\label{sec:moti}

\begin{figure}[t]
    \centering
    \includegraphics[width=0.85\linewidth]{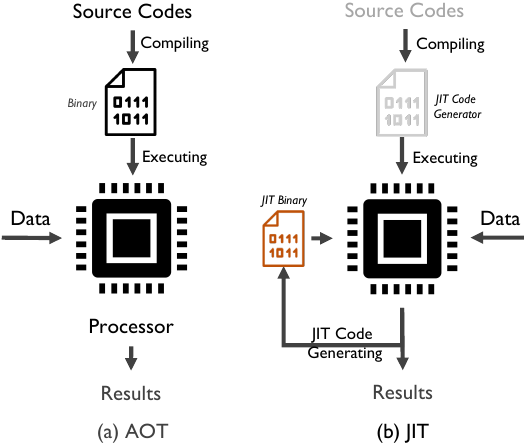}
    \caption{The difference between ahead-of-time (AOT) and just-in-time (JIT) code generation.}
    \label{fig:aotjit}
\end{figure}

\subsection{Just-in-time Assembly Code Generation}
Most existing solutions for SpMM computation adopt an \emph{ahead-of-time (AOT) compilation} approach\cite{merrill2016merge,wang2019deep,fey2019fast}, as depicted in \cref{fig:aotjit}(a). 
This process begins with a source code implementation of general SpMM computation written in conventional languages like C/C++. 
The next step is to compile the source code into an executable binary using mainstream compilers, such as \texttt{gcc}. 
Finally, the binary is executed with input data to compute the results.

Unfortunately, despite the meticulous optimization efforts invested in the source code implementation of SpMM\cite{wang2014intel,huang2020ge}, the AOT approach falls short of generating the most efficient executable binary, primarily due to three key factors:
1) C/C++ compilers rely on heuristic rule-based register allocation schemes\cite{chaitin1982register} that are inadequate at capturing the memory access pattern characteristics of SpMM computations, which lead to unnecessary accesses to memory.
2) An inherent limitation of the AOT approach is its inability to leverage runtime information, which inadvertently results in the introduction of additional branch instructions to handle varying input data. 
This can subsequently lead to overhead due to branch misprediction\cite{inoue2014faster}.
3) Redundant assembly instructions are introduced due to unnecessary memory accesses (e.g., register spill and memory loads) and branch control operations (e.g., comparison and conditional jump), leading to an excess of executed instructions.

In this work, we propose the adoption of \emph{just-in-time (JIT) assembly code generation}\cite{shaylor2002just,arnold2005survey} for SpMM, which follows a different process as shown in \cref{fig:aotjit}(b).
It starts with the development of a JIT code generator, which does not generate ``general purpose'' SpMM code.
Instead, when data is available at runtime (e.g., dimensions of the matrices), the JIT code generator can produce assembly code that is tailored to the specific instance of SpMM being executed.
In contrast to AOT solutions, this methodology can lead to a improved utilization of registers, decreased branch overhead, and reduction in the number of instructions, as demonstrated in next subsection.

\begin{table}[t]
    \centering
    \caption{Comparison of JIT and AOT compilation for single-thread scalar SpMM on the uk-2005 sparse matrix.}
    \begin{tabular}{r|cccc}
    \toprule
        & \textbf{Gcc} & \textbf{Clang} & \textbf{ICC} & \textbf{JIT} \cr \midrule
        \textbf{Execution Time (s)} & 8.6 & 9.1 & 6.3 & \emph{3} \cr
        \textbf{Memory Loads (billions)} & 2.2 & 2.3 & 2.4 & \emph{0.9} \cr
        \textbf{Branches (millions)} & 813 & 489 & 233 & \emph{196} \cr
        \textbf{Branch Misses (millions)} & 6.6 & 5.3 & 5.5 & \emph{2.7} \cr
         \textbf{Instructions (billions)} & 7.0 & 6.4 & 5.4 & \emph{1.6} \cr
    \bottomrule
    \end{tabular}
    \label{tab:scalar}
\end{table}

\subsection{Single-Thread Scalar SpMM Comparison}
\label{sec:jitVSAOTSingleThread}
To demonstrate the advantages of JIT assembly code generation over AOT approaches, we compile the sequential C implementation of SpMM in Algorithm \ref{alg:seqcsr} using three mainstream C compilers (\texttt{gcc}, \texttt{clang}, and \texttt{icc}\footnote{The versions of compilers used in this work are \texttt{gcc} 11.3.1, \texttt{clang} 15.0.7, and \texttt{icc} 2021.8.0.20221119. To ensure optimized compilation, the -O3 flag was employed for the compilation process.}) without using SIMD instruction or multi-threading.
We then compare their performance with a single-thread JIT implementation, which is a simplified version of our proposed solution (see \cref{sec:jitcodegen}) without using SIMD instructions.
We run all the implementations on the \emph{uk-2005} sparse matrix dataset\footnote{\url{https://sparse.tamu.edu/LAW/uk-2005}}, which is a square matrix with 39 million rows and 936 million non-zeros.
This sparse matrix is multiplied by a random-value dense matrix with 39 million rows and 8 columns.
The results are shown in \cref{tab:scalar}.
One can observe that even with the code generation overhead included, the execution time of the JIT approach is 2.1 -- 3\texttimes ~faster than the three AOT solutions.

The profiling data presented in \cref{tab:scalar} provides evidence that JIT assembly code generation effectively mitigates the three limitations associated with AOT solutions.
1) The JIT solution efficiently utilizes registers to store frequently accessed data, reducing the number of memory loads by 2.4 -- 2.7\texttimes ~when compared to AOT compilation.
For example, intermediate results of $ret$ from Algorithm \ref{alg:seqcsr} are stored in scalar registers \emph{XMM0-XMM7}, and the value of $A.vals[idx]$ at line 6 is stored in register \emph{XMM31}.
This allows the JIT program to perform read and write operations on these variables without accessing memory multiple times.
2) With the availability of runtime information that the number of columns $d$ in the dense matrix is 8, the JIT solution is able to unroll the for-loop at line 2 of Algorithm \ref{alg:seqcsr}.
As a result, the JIT solution reduces the number of branch instructions issued at the end of each iteration, which reduces the number of branch misses by 1.2 -- 4.1\texttimes ~compared to the AOT approaches.
Minimizing branch mispredictions is essential, as they can significantly degrade overall performance by causing pipeline flushes and refills\cite{farcy1998dataflow}.
3) By removing instructions related to unnecessary memory access and branch control, the JIT solution achieves the same computation with a reduction of 3.4 -- 4.4\texttimes~in the number of executed instructions.

\section{{\name} Framework}
\label{sec:jitcodegen}
\begin{figure}[t]
    \centering
    \includegraphics[width=0.75\linewidth]{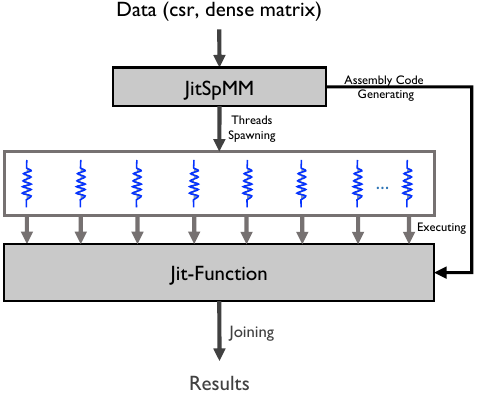}
    \caption{Overview of {\name}.}
    \label{fig:jitspmm}
\end{figure}

\subsection{Overview}
In \Cref{sec:jitVSAOTSingleThread}, we demonstrated significant advantages of the JIT assembly code generation technique in improving the performance of single-thread scalar SpMM.
However, there are several challenges in applying this technique to modern multi-threaded CPUs with SIMD instruction extensions.
The first challenge lies in effectively dividing the workload of the SpMM operation among the available CPU threads, without causing load imbalances that could hinder performance.
The second challenge involves scheduling the generated instructions to fully exploit the instruction-level parallelism of each CPU core. 
Last, efficiently mapping computations to SIMD instructions and utilizing SIMD registers requires careful consideration and optimizations.

To address these challenges, we propose {\name}, a just-in-time assembly code generation framework designed for high-performance SpMM computation on multi-core CPUs with SIMD instruction extensions. The workflow of {\name} is depicted in \cref{fig:jitspmm}.
With the input data of sparse and dense matrices, our {\name} framework follows a three-step process. 
First, it generates the assembly code for the computation and encapsulates it as a \emph{jit-function}. 
Then, a specific number of threads are created based on the hardware configuration.
Each thread independently determines its assigned workload and invokes the \emph{jit-function} to perform the computation on its assigned portion of the data.
Once all threads complete their respective workloads, they synchronize and merge into a single thread. 
Finally, the merged thread consolidates the individual results and returns the final output of the computation.
In the following subsections, the detail of our {\name} framework is presented in three parts, each addressing a specific challenge previously identified.
First we address the challenge of workload division through the integration of the JIT assembly code generation technique into three commonly employed workload division methods.
Next we present our \emph{coarse-grain column merging} technique to maximize the instruction-level parallelism when scheduling the generated instructions in the \emph{jit-function}.
Last, we demonstrate how {\name} leverages SIMD registers and instructions to minimize memory accesses and maximize arithmetic throughput.

\subsection{Workload Division}
\begin{figure}[t]
    \centering
    \includegraphics[width=0.85\linewidth]{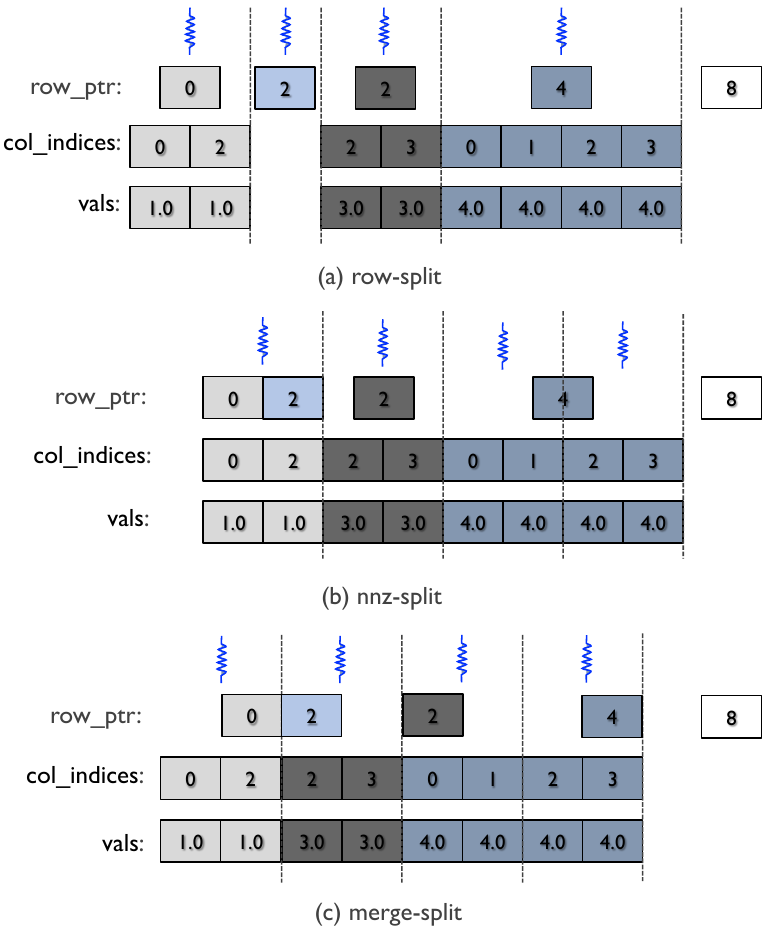}
    \caption{Three different workload assignment strategies for the thread-level parallelism.}
    \label{fig:tlp}
\end{figure}

In this subsection, we provide a concise overview of the workload division methods employed in {\name} and illustrate the integration of our JIT assembly code generation technique into these methods.

\subsubsection{Row-split, nnz-split and merge-split} Due to the irregular distribution of non-zero elements across various rows within the input sparse matrix, attaining efficient workload distribution is essential for achieving good SpMM performance\cite{merrill2016merge}. In the context of {\name}, we encompass three widely-utilized strategies\cite{yang2018design,merrill2016merge}: \emph{row-split}, \emph{nnz-split}, and \emph{merge-split}, as depicted in \cref{fig:tlp}. 
Each of these three methods selects a specific object to distribute evenly among all CPU threads.
Specifically, the \emph{row-split} approach evenly assigns rows of the sparse matrix to each thread, as illustrated in \cref{fig:tlp}(a).
Conversely, the \emph{nnz-split} method distributes an identical number of non-zero elements to each thread, as depicted in \cref{fig:tlp}(b).
The \emph{merge-split} strategy aims to equalize the total count of rows and non-zero elements across threads, as showcased in \cref{fig:tlp}(c).

It's worth noting that the \emph{row-split} strategy may result in workload imbalance, as specific threads might receive a significantly higher or lower number of non-zero elements\cite{fu2022tlpgnn}.
Consider \cref{fig:tlp}(a), where the second thread does not receive any non-zero elements, while the fourth thread is assigned four of them.
To address this issue, we introduce a \emph{dynamic row dispatching} method based on the \emph{row-split} method.
\thomas{What is ``it'' at the end of the above sentence?}\qiang{Done.}
Listing \ref{lst:drd} provides an x86 implementation of dynamic row dispatching for each thread in the SpMM computation.
First, a global integer variable named \texttt{NEXT} is allocated to keep track of the next unprocessed row ID. 
It is initialized to zero before the computation begins.
Each thread requests a batch of rows dynamically by atomically reading the value of \texttt{NEXT} and adding the batch size\footnote{The batch size is set to 128 in this work.} to it via the \texttt{xadd} instruction\cite{x86xadd2023}.

\begin{lstlisting}[style=x86asm, caption={Assembly implementation of dynamic row dispatching for one thread.}, label=lst:drd]
    ; load the address of NEXT before the loop
    mov rdi, #address_of_NEXT
.start:
    ; load the batch number
    mov rsi, #batch_number
    ; atomic exchange and add
    lock xadd QWORD PTR [rdi], rsi
    ; boundry check
    cmp rsi, #nrow
    jge .end
    ... ; instructions for computation of one batch
    jmp .start
.end:
    ret
\end{lstlisting}

\subsubsection{Integration of JIT} Considering that the AOT implementations of these three workload assignment methods are also susceptible to the limitations highlighted in \cref{sec:moti}, we employ our JIT code generation technique by generating different assembly codes for each approach. 
Each of the three strategies has distinct workload distributions assigned to each thread. 
As a result, they require different implementations of the computations performed by the threads.
Our {\name}'s code generation process begins with the assembly codes for computing a single row, which remains the same across all three methods.
For the \emph{row-split} strategy with dynamic row dispatch, we incorporate the code snippets in Listing \ref{lst:drd} for workload requesting. 
These instructions facilitate the dynamic allocation of work to each thread.
For the \emph{nnz-split} and \emph{merge-split} approaches, we introduce additional instructions to implement loops that execute the computations on a continuous range of rows, which is determined by a binary search employed by the two approaches\cite{merrill2016merge}. 
This allows the threads to perform matrix multiplication on the assigned segments.
It is worth noting that our JIT code generation framework enables the seamless integration of these different workload assignment techniques. 
By dynamically generating tailored assembly codes, we optimize the SpMM computations for each strategy, ensuring efficient utilization of computational resources and achieving high-performance parallel processing.
In the next two subsections, our attention shifts to the generation of optimized code for computing a single row of the sparse matrix, specifically targeting lines 2-7 of Algorithm \ref{alg:seqcsr}.

\subsection{Coarse-grain Column Merging}

In this subsection, we introduce our \emph{coarse-grain column merging} (CCM) technique, which aims to maximize instruction-level parallelism (ILP) of the assembly codes generated for the computation of a single matrix row, with the help of runtime information.
By enabling the simultaneous execution of multiple instructions, ILP helps keep the processor's computing resources, such as the pipeline and floating-point units (FPUs), busy\cite{pai1997impact,huang2020ge}.

\subsubsection{High-level Overview}

\begin{algorithm}[t]
\footnotesize
\caption{Coarse-grain Column Merging (CCM) for computing a single row in SpMM with $d=45$}\label{alg:ccm}
\SetKwInOut{Input}{Input}
\SetKwInOut{Out}{Output}

\Input{Row index $i$, $A.row\_ptr$, $A.col\_indices$, $A.vals$, and dense matrix $X$.}
\Out{Result of the $i$-th row $Y[i][0\colon 45]$.}
    \SetInd{0.1em}{0.6em}
    $ret[0\colon 45] := 0$ \tcp{All elements initialized to zero}
    \For{$idx = A.row\_ptr[i]\ \textbf{to}\ A.row\_ptr[i+1]$} {
        $k = A.col\_indices[idx]$ \\
        $ret[0\colon 45] \mathrel{+}= A.vals[idx] * X[k][0\colon 45]$ \tcp{Mul-and-add}
    }
    $Y[i][0\colon 45] = ret[0\colon 45]$ \tcp{Write result back}
\end{algorithm}

The CCM technique is designed to optimize a single row SpMM computation by unrolling the for-loop at line 2 in Algorithm \ref{alg:seqcsr} and merging all the columns of an entire row as a single vector, given the runtime information of columns $d=45$ of the dense matrix $X$ as example.
Algorithm \ref{alg:ccm}, which will be used to replace lines 2-7 of Algorithm \ref{alg:seqcsr} at runtime, outlines the steps involved in implementing CCM.
First, we allocate and initialize a vector of length $45$, i.e., $ret[0\colon 45]$, to store the $45$ elements of the resulting row, as shown at Line 1, where $:=$ means assigning a scalar value to every single element in a vector.
Next, we iterate over the non-zero (nz) element list of row $i$ (Line 2).
For each nz element, we load its value and multiple it with each element in the corresponding row in the input dense matrix $X[k][0\colon 45]$. 
The resulting vector is added to the accumulating vector $ret[0\colon 45]$ (Line 4).
Finally, the accumulated vector is written back to row $i$ of the resulting matrix $Y$ (Line 5).
It is essential to emphasize that CCM becomes viable only when the runtime information about the column number $d=45$ of the input dense matrix is known. 
In contrast, all AOT solutions are constrained to use a loop ranging from 0 to $d$, thus missing out on the benefits elaborated in the subsequent paragraph.
\thomas{I am having some trouble understanding the technique here, as presented in the algorithm. On line 4, it looks like the left-hand-side $v$ is a vector of length $d$, but is initialized/assigned a single scalar value, namely the non-zero value from $A$ at index $idx$. I would have assumed that the point here was to construct a full vector $v$ of $d$ elements; what does it mean to broadcast to a vector here? Is there some assumed SIMD loading happening? I am also having some trouble seeing how Algorithm 2 relates to the unrolling of the loop from Algorithm 1; where does the ``code'' in Algorithm 2 go/fit into Algorithm 1? Does Algorithm 2 replace the code in Algorithm 1 on lines 3-7? Also, what is Algorithm 2 returning and to where? Is it the entire row $i$ in the output matrix $Y$? I think the usual notation for that, since $Y$ is a 2-d array/matrix, is something like $Y[i,:]$, kind of like python. Or I suppose you would use $Y[i, 0:d]$ like you have on line 6. So yeah, why return $Y[i]$ and what does that represent?}\qiang{Rewrote this paragraph, hope can resolve all you confusions.}

\subsubsection{Benefits}

The CCM technique offers several advantages that enhance ILP. 
First, it enables independent workload processing. 
Since the computations for different column indices are independent, the instructions used for single vector computations (such as lines 1, 4, or 5 in Algorithm \ref{alg:ccm}) are also independent. 
This lack of data dependence among instructions eliminates possible pipeline stalls and promotes better ILP.
Second, CCM eliminates branch overhead by unrolling the original loop on line 2 in \ref{alg:seqcsr} that iterates over each column index, which is enabled by knowing the column number $d$ at runtime.
This allows us to avoid executing the branch instruction associated with the loop. 
\thomas{Which branch instruction is this referring to? Are we talking about the branch instruction that would be executed for each iteration of the outer loop that iterates over $m$? And in case I forgot, or it wasn't mentioned, are we unrolling the loop over $d$ completely? If not, what is the unrolling factor? If we are unrolling it entirely, then yes the outer loop over $m$ would not pay the penalty of $d$ branch instructions for the loop over $d$. But if we are not unrolling completely, there would still be $d/F$ branch instructions, where $F$ is the unrolling factor.}\qiang{There is not change on the outer loop. The eliminated branch instruction are used for the loop over column $0$ to $d$. And yes, we are unrolling it entirely.}
Moreover, this eliminates the penalty associated with branch mispredictions, which can significantly hinder pipeline efficiency and degrade ILP due to the pipeline flushing and instruction refilling\cite{eyerman2006characterizing}.
The last advantage lies in the improved memory access pattern. 
During the SpMM computation, a substantial amount of memory operations involve loading the corresponding row in the dense input matrix, denoted as $X[k][0\colon 45]$, for each nz element, as depicted in Line 5 of Algorithm \ref{alg:ccm}.
Without coarse-grain column merging, the program would need to iterate over all column indices. 
In each iteration, it reads values at the same column index for all corresponding rows in the dense matrix before moving on to the next column index.
\begin{figure}[t]
    \centering
    \includegraphics[width=0.85\linewidth]{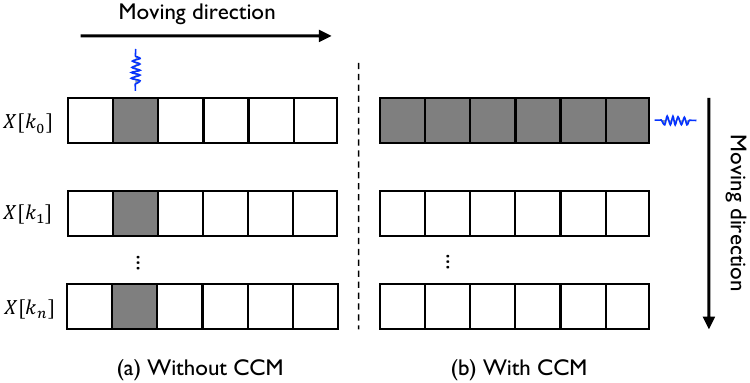}
    \caption{Comparison between the memory access patterns with and without our \emph{coarse-grain column merging (CCM)} technique.}
    \label{fig:mem}
\end{figure}
This results in non-sequential memory accesses since values at the same column index of different rows are not stored in contiguous memory addresses (as shown in \cref{fig:mem}(a)), considering the row-major format of the matrix.
In contrast, our coarse-grain column merging technique processes all the columns of a single row in one step, as illustrated in \cref{fig:mem}(b). 
This enables sequential memory accesses as the values are stored consecutively in memory. 
Consequently, this leads to a reduction in cache misses, minimizes pipeline stalls for memory loads, and ultimately enhances ILP\cite{hennessy2011computer}.

\subsection{Register Allocation and Instruction Selection}

In this subsection, we delve into the generated assembly codes of our {\name} framework to demonstrate how we leverage SIMD registers and instructions to reduce memory accesses and improve arithmetic throughput.
Listing \ref{lst:gen} provides an example of the assembly code generated for the computation of a single row, where the number of columns is set to 45 and the value type is 32-bit floating-point. 

\subsubsection{Register Allocation}
In the x86 architecture, there are three classes of SIMD registers: \emph{XMMs}, \emph{YMMs}, and \emph{ZMMs}. These registers have sizes of 128 bits, 256 bits, and 512 bits, respectively. This means that they can hold up to 4, 8, and 16 single-precision floating-point values.
The low-order 256 bits of the \emph{ZMM} register are aliased to a corresponding \emph{YMM} register, and the low-order 128 bits of the \emph{YMM} register are aliased to a corresponding \emph{XMM} register. 
This allows for compatibility and flexibility in using different register sizes.
For CPUs that support the AVX512 instruction extension, there are a total of 32 \emph{ZMM/YMM/XMM} registers available. 
These registers provide a significant amount of storage for parallel computation and can greatly enhance the performance of SIMD operations.
It is worth mentioning that scalar floating-point operation instructions can also use the \emph{XMM} registers as operands. 
In this case, only the low-order 32 or 64 bits of these registers are utilized, which is suitable for scalar operations that do not require SIMD parallelism.

\begin{figure}[t]
    \centering
    \includegraphics[width=0.9\linewidth]{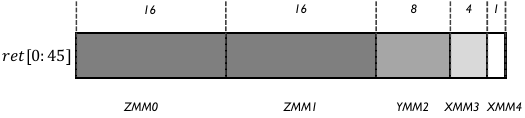}
    \caption{The register allocation for storing the vector $ret[0\colon d]$ of single row SpMM computation with $d=45$ and 32-bit floating-point number type.}
    \label{fig:reg}
\end{figure}

Our objective in register allocation is to reduce memory accesses by improving the retention of data in registers.
To achieve this, we employ a combination of SIMD registers to store the entire vector $ret$ in Algorithm \ref{alg:ccm}.
To handle arbitrary sizes of the vector, we decompose the size into a linear combination of sizes that can be stored in different types of SIMD registers (\emph{ZMM/YMM/XMM}), while using the fewest number of registers possible. 
For example, consider the case of $d=45$ with 32-bit floating-point values. 
We break it down as $16(\emph{ZMM0})+16(\emph{ZMM1})+8(\emph{YMM2})+4(\emph{XMM3})+1(\emph{XMM4})$. 
The last register (\emph{XMM4}) is used for storing a scalar value.
By storing the vector $ret[0\colon 45]$ in SIMD registers, we can avoid the need for memory accesses during the update process (line 5 of Algorithm \ref{alg:ccm}).
Additionally, we broadcast the nz value $A.vals[idx]$ in Line 4 into register \emph{ZMM31}, eliminating the need for multiple memory loads.
\thomas{Going back to my confusion from Algorithm 2, we are broadcasting a single scalar into zmm31, right? How/where are we packing multiple values into the SIMD register? The FMA ops after this broadcast go through the high and low order bits of the zmm register, but didn't we only put a single scalar (vals[r12]) into it?}\qiang{By broadcasting a scalar value (vals[r12]) into the zmm31 register. Each of the 16 32-bits float-point values in zmm31 is equal to vals[r12]. And following vfmadd231ps instruction use this value for the vector fused mul and add computation.}
To keep track of the row's nz list and its corresponding indices, we utilize three general-purpose registers (\emph{r10}, \emph{r11}, and \emph{r12}) to store the start and end positions of the row's nz list, as well as the corresponding row index ($A.row\_ptr[i]$, $A.row\_ptr[i+1]$, and $A.col\_indices[idx]$).
Strategically allocating registers allows us to retain data within the registers for extended periods, reducing the need for frequent and costly memory access operations. 
Consequently, the program can execute computations using data stored in the registers, leading to faster processing times and improved overall performance.
\subsubsection{Instruction Selection}

\begin{lstlisting}[style=x86asm, caption={Generated assembly code for computing a row $i$ of SpMM with column number of 45 for the dense matrix and 32-bit floating-point number as the value type.}, label=lst:gen]
    mov rdi, #i ; load the row index to be processed
    ; initialize the registers storing the results
    vxorps zmm0, zmm0, zmm0
    vxorps zmm1, zmm1, zmm1
    vxorps ymm2, ymm2, ymm2
    vxorps xmm3, xmm3, xmm3
    vxorps xmm4, xmm4, xmm4
    ; load the start and end position of the nz list
    mov r10, #row_ptr[rdi]
    mov r11, #row_ptr[rdi+1]
.nnzloop_start: ; loop over the nz list
    ; boundry check
    cmp r10, r11
    jge .nnzloop_end
    ; load corresponding row id
    mov r12, #col_indices[r10]
    ; load the nz value and broadcast it to zmm31
    vbroadcastss zmm31, #vals[r12]
    ; accumulate the results
    vfmadd231ps zmm0, zmm31, #X[r12,0:16]
    vfmadd231ps zmm1, zmm31, #X[r12,16:32]
    vfmadd231ps ymm2, ymm31, #X[r12,32:40]
    vfmadd231ps xmm3, xmm31, #X[r12,40:44]
    vfmadd231ss xmm4, xmm31, #X[r12,44]
    ; next nz element
    inc r10
    jmp .nnzloop_start
.nnzloop_end:
    ; write the result into memory
    vmovups #Y[rdi,0:16], zmm0
    vmovups #Y[rdi,16:32], zmm1
    vmovups #Y[rdi,32:40], ymm2
    vmovups #Y[rdi,40:44], xmm3
    vmovss #Y[rdi,44], xmm4
\end{lstlisting}
The x86 architecture's SIMD extension offers programmers a wide range of instructions to perform various operations on multiple data items simultaneously, including arithmetic computations, logical operations, data shuffling, and data movement\cite{x86manual2023}. 
When generating assembly codes for the computation of a single row, we utilize three key classes of SIMD instructions. 
First, we employ packed floating-point bit-wise exclusive OR (\texttt{vxorps}) instructions to zero the registers representing the result vector $ret[0\colon 45]$, as demonstrated in lines 3--6 of Listing \ref{lst:gen}. 
These instructions are preferable to data movement instructions like \texttt{vmovups} as they avoid modifying the program status and control register\cite{kusswurm2014modern}. 
Second, we utilize packed fused-multiply-add (FMA) instructions (\texttt{vfmadd231ps}) to accumulate the multiplication results of the nz value and the corresponding row in the dense matrix (Line 20--23).
FMA instructions are capable of performing both floating-point multiplication and addition with a single rounding operation, resulting in higher arithmetic throughput compared to two separate instructions\cite{kusswurm2014modern}.
Last, we employ packed floating-point movement instructions (\texttt{vmovups}) to store the result back into memory (Line 30--33).
In summary, by leveraging these SIMD instructions, we are able to harness the power of data parallelism in our generated binary. 

\section{Experiment and Evaluation}
In this section, we showcase the results of our {\name} performance comparison against baseline solutions. 
Furthermore, we delve into a comprehensive analysis of the performance using profiling data to gain deeper insights.
\thomas{We probably should add a short paragraph here that says what we're going to show. Something like ``To demonstrate the performance advantages of our {\name} framework, we conducted an in depth performance evaluation.'' Then you can describe what you're going to show (e.g., comparison to auto-vectorization and Intel MKL). Also mention that you perform profiling/analysis to better understand the performance. Basically, this paragraph is just a summary of what this section will present.}\qiang{Done.}
\label{sec:exp}

\begin{table}[t]
    \centering
    \caption{Statistics of sparse matrix datasets.}
    \begin{tabular}{l|rr}
    \toprule
    \textbf{Name} & \textbf{rows} & \textbf{nnz} \cr \midrule
    mycielskian19 &	393,215 & 903,194,710 \cr
    uk-2005 & 39,459,925 & 936,364,282 \cr
    webbase-2001 & 118,142,155 & 1,019,903,190 \cr
    it-2004 & 41,291,594 & 1,150,725,436 \cr
    GAP-twitter & 61,578,415 & 1,468,364,884 \cr
    twitter7 & 41,652,230 & 1,468,365,182 \cr
    GAP-web & 50,636,151 & 1,930,292,948 \cr
    sk-2005 & 50,636,154 & 1,949,412,601 \cr
    mycielskian20 & 786,431 & 2,710,370,560 \cr
    com-Friendster & 65,608,366 & 3,612,134,270 \cr
    GAP-kron & 134,217,726 & 4,223,264,644 \cr
    GAP-urand & 134,217,728 & 4,294,966,740 \cr
    MOLIERE\_2016 & 30,239,687 & 6,677,301,366 \cr
    AGATHA\_2015 & 183,964,077 & 11,588,725,964 \cr
    \bottomrule
    \end{tabular}
    \label{tab:dataset}
\end{table}

\subsection{Experiment Setup}
\subsubsection{Datasets and Environment}
To evaluate the performance of our {\name} framework on large input matrices, we conducted experiments using the 14 largest sparse matrices (in terms of number of nonzero elements) from the SuiteSparse Matrix Collection\cite{davis2011university}.
The statistics of these datasets are presented in \cref{tab:dataset}, where all matrices are square.
\thomas{I changed the table to just say ``rows'' since having ``nrow/ncol'' may seem like ``number of rows divided by number of columns'' to some people who may not realize that what we mean is that the matrices are square and nrow == ncol.}\qiang{Thanks.}
We generated the dense input matrix consisting of random 32-bit floating-point values and either 16 or 32 columns (the number of rows is equal to the number of columns of the sparse input matrix).
\thomas{Are the results we present only with column numbers 16 and 32? Assuming that is true, I specified that above.}\qiang{Thanks.}
The implementation of our {\name} framework was developed in C/C++ with OpenMP\cite{chandra2001parallel} and comprised approximately 2,000 lines of code. 
To generate x86 assembly code at runtime, we utilize the open-source framework AsmJit\cite{AsmJit2023}.
All experiments were conducted on a server equipped with 24-core Intel Xeon(R) Gold 6126 CPU and 1.5 TB of DRAM.
All implementations were executed using 48 threads, taking advantage of hyper-threading.
\thomas{Are all the experiments in this section multi-threaded? If so, how many threads are utilized? I assume 24 given we have a 24-core CPU? Worth mentioning it either way.}\qiang{Done.}
The server runs a Centos 8 system with kernel version of 4.18.0 and GCC version 11.3.1.
To ensure reliable results, we performed each experiment ten times and report the average values.
\thomas{Do we have a variation metric we can report? What I usually do is report the coefficient of variation between trials: compute the standard deviation and the mean of the trials. Then divided the standard deviation by the mean. You want to see a small value (e.g., 0.001), which indicates very little variation. If you have a larger number, like 0.5, it means a lot of variation. A value of 1 would mean that your standard deviation is equal to your mean, which is bad! In any case, all you need to say here is: ``The variation between trials did not exceed X\%'', where X is the largest coefficient of variation you found across all the experiments, multiplied by 100 to get a percentage.}\qiang{Good point. But we don't have time to collect this number now.}
The profiling results are collected using Linux Perf\cite{de2010new}.

\subsubsection{Baselines Solutions}
In the subsequent subsections, we provide a comprehensive analysis of the results obtained from comparing our {\name} framework with two baseline AOT solutions: auto-vectorization and Intel's Math Kernel Library (MKL).
All mainstream C/C++ compilers offer \emph{auto-vectorization}, a optimization technique that analyzes the code to identify loops and computations suitable for vectorization. 
If such loops are found, SIMD instructions that are compatible with the processor's vector units (e.g., AVX-512) are generated automatically by the compiler.
To evaluate the performance of our {\name} framework against auto-vectorization, we implemented the three different workload assignment methods for SpMM in pure C/C++ by extending the implementations presented in Merrill and Garland's work\cite{merrill2016merge} and compiling them using the Intel \texttt{icc} compiler with the \texttt{-O3 -mavx512f} flags to enable vectorization with AVX-512 instructions\footnote{We chose the \texttt{icc} compiler because \texttt{gcc} and \texttt{clang} did not generate AVX-512 instructions in the output binary for the SpMM implementations.}.
\thomas{For the footnote above, does this mean that gcc and clang CANNOT generate AVX-512, or they did not for our particular use-case for some reason?}\qiang{They can generate AVX512 for simple input such as a single for loop.}
The Intel MKL\cite{wang2014intel} is a widely used collection of highly-optimized functions and routines designed to accelerate a comprehensive set of mathematical functions that encompass various domains, including linear algebra, fast Fourier transforms (FFTs), sparse matrix operations, and statistical analysis.
Functions and routines within MKL are hand-crafted through low-level coding, including C/C++ and assembly, with adaption of SIMD vectorization and thread parallelism.
We use the \texttt{mkl\_sparse\_spmm} routine from Intel MKL as our another baseline.
\thomas{Is there a specific MKL routine that we use to compare against?}\qiang{Done.}

\subsection{Code Generation Overhead}

\begin{table}[t]
    \centering
    \caption{The execution (Exe) time (in seconds) and the code generation (Codegen) overhead (\%) of {\name} with \emph{row-split} workload assignment method with column number of 16.}
    \begin{tabular}{l *{2}{d{3.3}} }
    \toprule
 \mcl{\textbf{Dataset}} & \mc{\textbf{Exe (s)}} & \mc{\textbf{Codegen Overhead (\%)}} \\ \midrule
    mycielskian19 &	0.43 & 0.0136\% \cr
    uk-2005 & 0.27 & 0.0217\% \cr
    webbase-2001 & 0.65 & 0.0090\% \cr
    it-2004 & 0.3 & 0.0201\% \cr
    GAP-twitter & 2.9  & 0.0028\% \cr
    twitter7 & 3.1 & 0.0020\% \cr
    GAP-web & 0.44 & 0.0138\% \cr
    sk-2005 & 0.43 & 0.0146\% \cr
    mycielskian20 & 2.03 & 0.0029\% \cr
    com-Friendster & 9.04 & 0.0007\% \cr
    GAP-kron & 9.51 & 0.0008\% \cr
    GAP-urand & 11.0 & 0.0007\% \cr
    MOLIERE\_2016 & 16.2 & 0.0004\% \cr
    AGATHA\_2015 & 22.5 & 0.0003\% \cr
    \bottomrule
    \end{tabular}
    \label{tab:codegen}
\end{table}

\begin{figure*}[t]
    \centering
    \subfigure[Column number of 16.]{\includegraphics[width=0.49\linewidth]{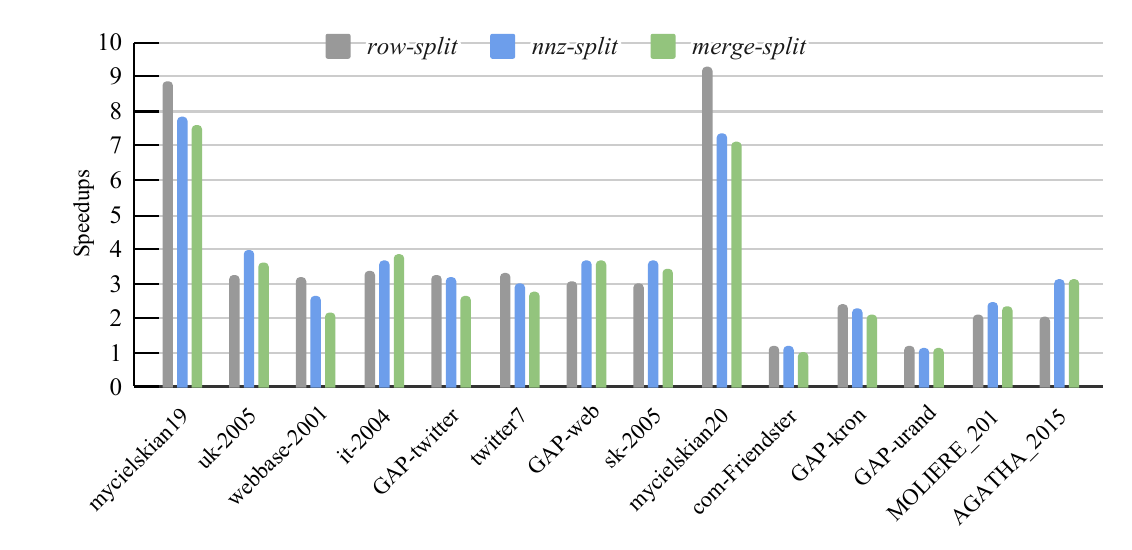}}
    \subfigure[Column number of 32.]{\includegraphics[width=0.49\linewidth]{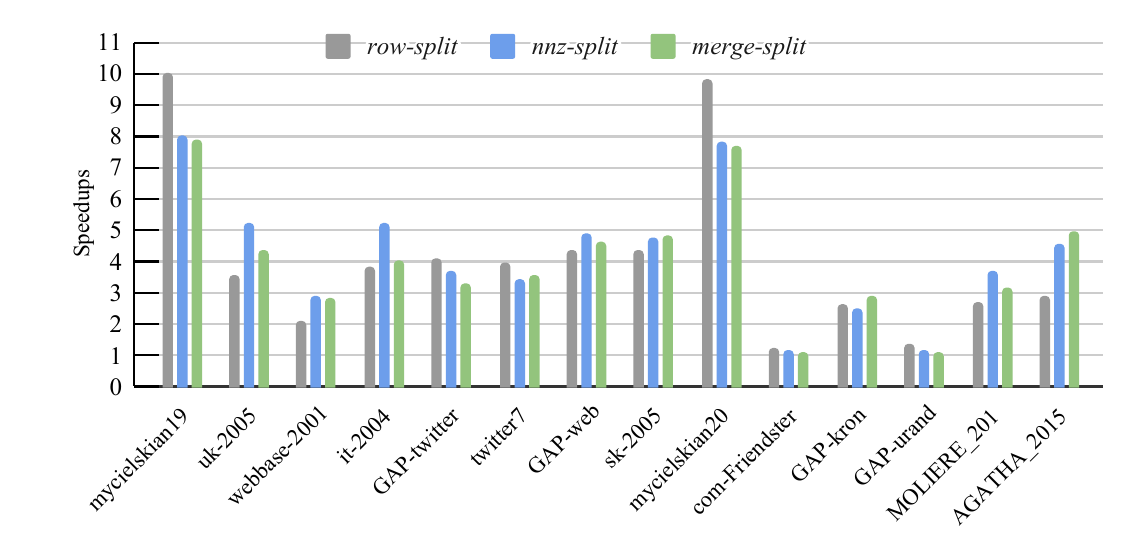}}
    \caption{Speedups of {\name} over Auto-Vectorizaiton solutions using Intel \texttt{icc} compiler with \texttt{-O3 -mavx512f} flags. Higher bars represent better performance.}
    \label{fig:jitvsauto}
\end{figure*}

\begin{figure*}[t]
    \centering
    \subfigure[Column number of 16.]{\includegraphics[width=0.49\linewidth]{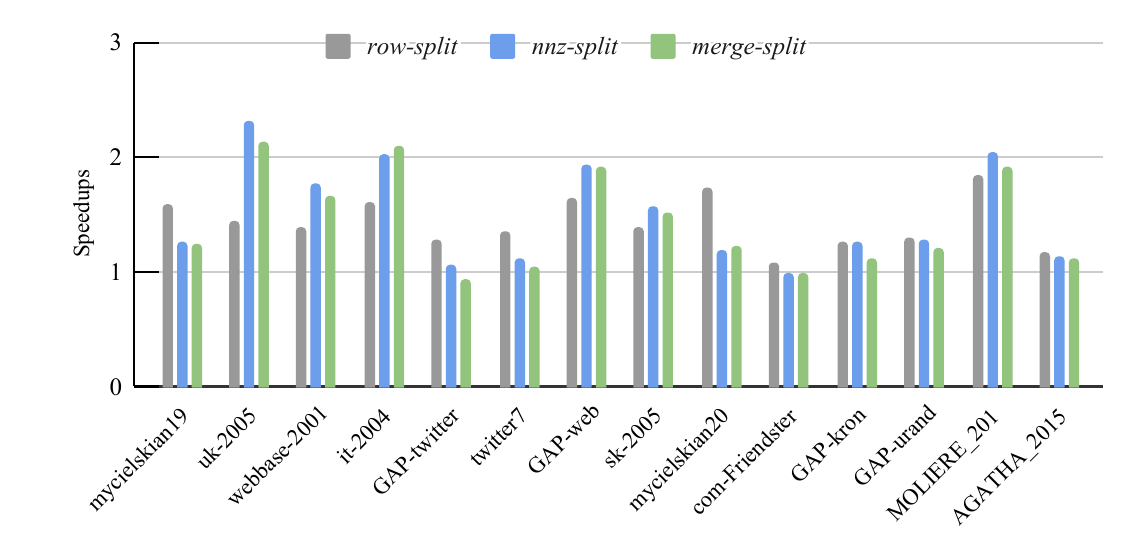}}
    \subfigure[Column number of 32.]{\includegraphics[width=0.49\linewidth]{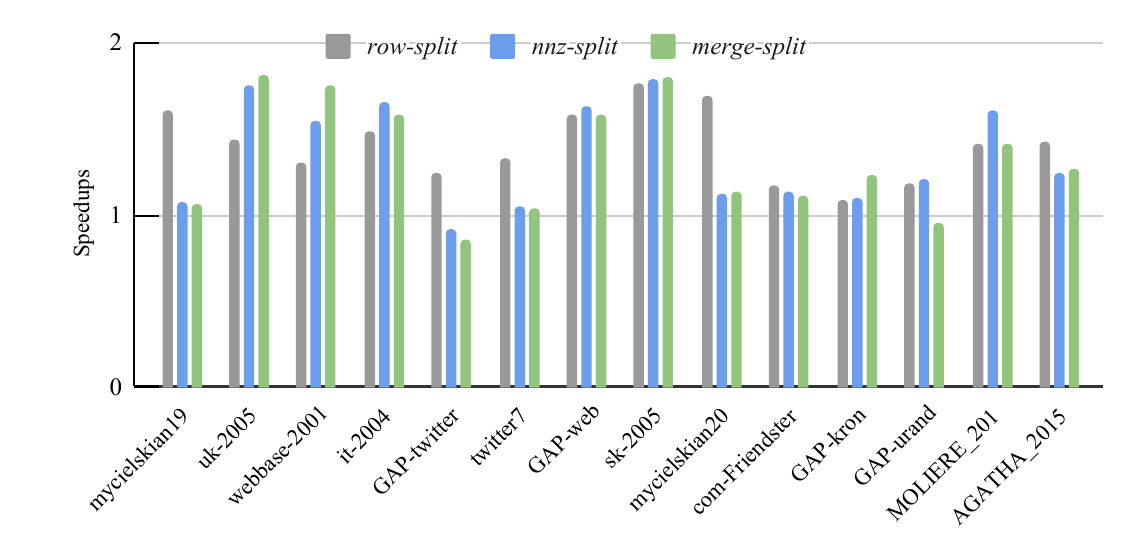}}
    \caption{Speedups of {\name} over Intel MKL. Higher bars represent better performance.}
    \label{fig:jitvsmkl}
\end{figure*}

Our {\name} involves generating assembly codes for SpMM computation and subsequently executing these generated codes as a function to obtain results, thus the execution time of {\name} encompasses the code generation overhead. 
\cref{tab:codegen} presents the execution time (in seconds) alongside the ratio of code generation time relative to the total execution time.
\thomas{I am wondering if we need to explicitly list the codegen runtimes in the table. I think showing the execution time and the codegen overhead as a percentage (as you do in the far right column) is enough; it gets across the point that the codgen overhead is negligible. So I'd suggest removing that middle column (codgen) and changing the name of the far right column to ``Codgen Overhead (\%)''.}\qiang{Agree. Here we just want to show the code generation overhead is negligible compared with overall execution time.}
For these results, we executed {\name} using the \emph{row-split} workload assignment method with column number of 16 for the input dense matrix. 
The results demonstrate that code generation times are minimal in comparison to total execution times, averaging 0.0074\% (down to 0.0003\%) of the total execution time. 
\thomas{What exactly do you mean in the last sentence? Are there other overheads in the codegen time reported other than the actual code generation? If not, then I think the last sentence here is a bit redundant (since the sentence before that says that code gen has minimal overhead). But if there are other overheads other than code generation, we should state them more clearly here.}\qiang{Agree.}

\subsection{Performance Comparison}
\subsubsection{Comparison with Auto-Vectorization}

For each of the three workload assignment methods, we compared the performance of our {\name} framework to the binary generated by \emph{icc}'s auto-vectorization. 
The speedups achieved by {\name} over auto-vectorization are illustrated in \cref{fig:jitvsauto}. 
{\name} outperforms the auto-vectorization solution across all workload assignment methods for different datasets and column numbers.
When considering a column number of 16, the average speedups are 3.5\texttimes~(up to 9.3\texttimes), 3.5\texttimes~(up to 7.8\texttimes), and 3.3\texttimes~(up to 7.6\texttimes) for \emph{row-split}, \emph{nnz-split}, and \emph{merge-split}, respectively. 
As column numbers increase to 32, the speedups become even more significant, with average values of 4.1\texttimes~(up to 10.0\texttimes), 4.2\texttimes~(up to 8.0\texttimes), and 4.1\texttimes~(up to 7.9\texttimes) for \emph{row-split}, \emph{nnz-split}, and \emph{merge-split}, respectively.
\thomas{Anything we can say here about why using a larger column number leads to larger performance gains? Is it just a matter of larger columns == more memory accesses, which we demonstrated earlier is something our approach aims to address? Maybe you mention this in Section D. If so, another reason why we should include a summary paragraph at the beginning of this section, so readers know to look for the profiling analysis.}\qiang{Yes. Should because larger columns means more memory access and branch control for AOT.}


\subsubsection{Comparison with Intel MKL}

We executed SpMM using Intel MKL on the 14 datasets in \cref{tab:dataset} with column numbers of both 16 and 32 for the input dense matrix, and compared those results to the {\name} framework.
\thomas{Again, which MKL routine did we use? May not need to mention it here if we also mention it earlier in the section.}\qiang{Mentioned before.}
The runtime speed-ups over MKL achieved by {\name} are presented in \cref{fig:jitvsmkl}. 
For datasets with a column number of 16, the \emph{row-split}, \emph{nnz-split}, and \emph{merge-split} approaches exhibit average speedups of 1.4\texttimes~(up to 1.9\texttimes), 1.5\texttimes~(up to 2.3\texttimes), and 1.4\texttimes~(up to 2.1\texttimes) respectively, outperforming Intel MKL's SpMM routine. 
Similarly, for column number of 32, the \emph{row-split}, \emph{nnz-split}, and \emph{merge-split} methods achieve average speedups of 1.4\texttimes~(up to 1.8\texttimes), 1.3\texttimes~(up to 1.8\texttimes), and 1.3\texttimes~(up to 1.8\texttimes) respectively, over Intel MKL.
\thomas{Again, anything else we can say here about the results, or will that be provided in Section D? This is mostly a note to myself.}\qiang{Since intel MKL is not open sourced, it is hard to put more explanations here.}

\subsubsection{Summary}
To summarize, the performance enhancements demonstrated in these experiments show the benefits of our {\name} framework over AOT compilation solutions.
These improvements are achieved when compared to auto-vectorization techniques (up to 10\texttimes) and hand-crafted libraries that are highly optimized (up to 2.3\texttimes).
In the next section, we will provide further insight into the performance of {\name} through a comprehensive profiling analysis.
\thomas{I think the above wording is a bit strong, so I changed some of it to come off ``softer''. It isn't to say that what you said isn't true, but you don't want a reviewer who has ton a bunch of work in AOT who thinks JIT is not that great to be ``offended'' here.}\qiang{Thanks.}

\subsection{Profiling Analysis}

\begin{figure*}[t]
    \centering
    \subfigure[Number of memory loads.]{\includegraphics[width=0.49\linewidth]{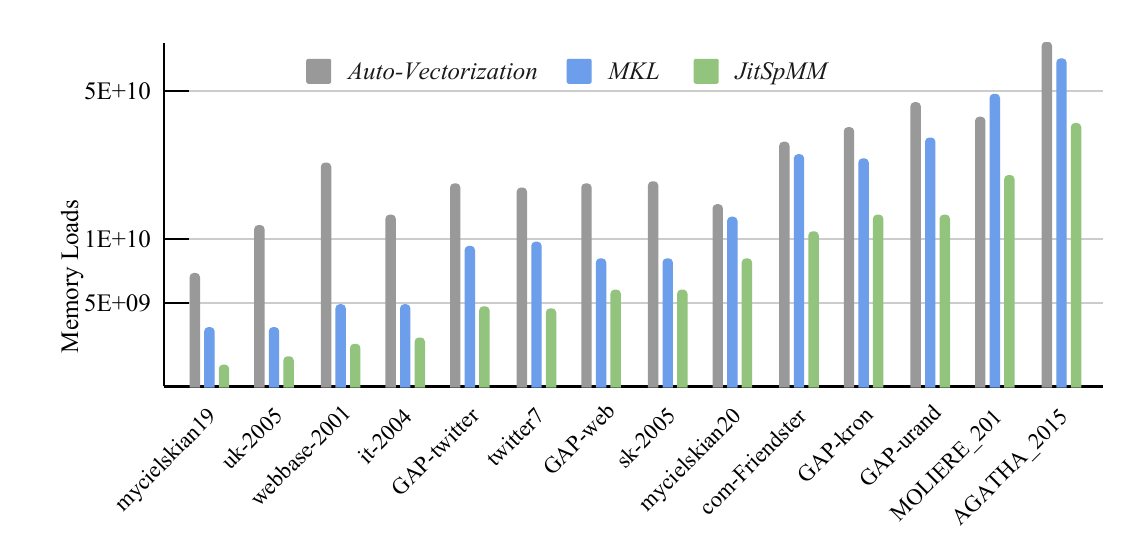}}
    \subfigure[Number of executed branch instructions.]{\includegraphics[width=0.49\linewidth]{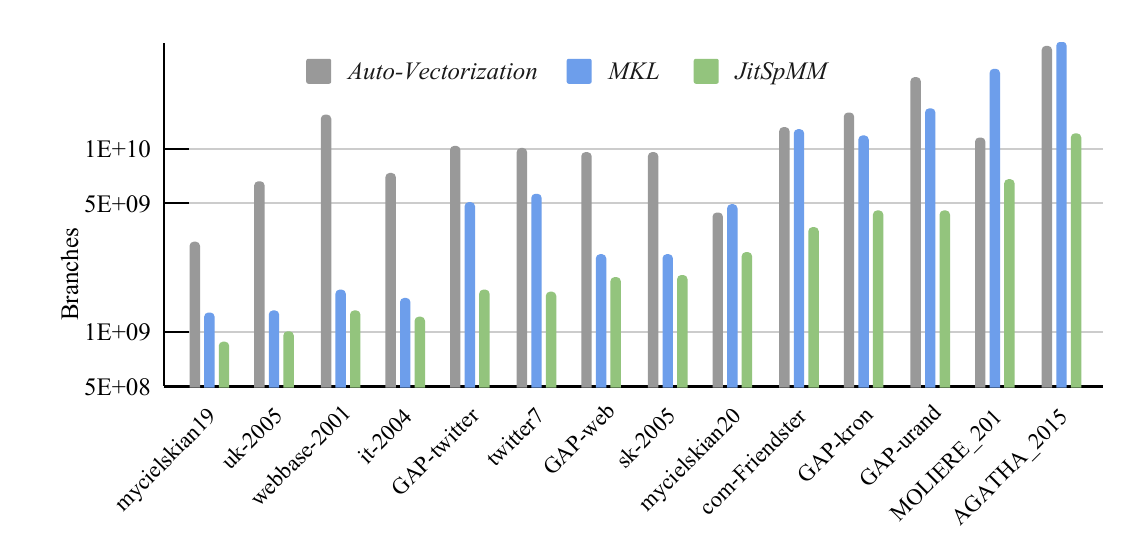}}
    \subfigure[Number of branch misses.]{\includegraphics[width=0.49\linewidth]{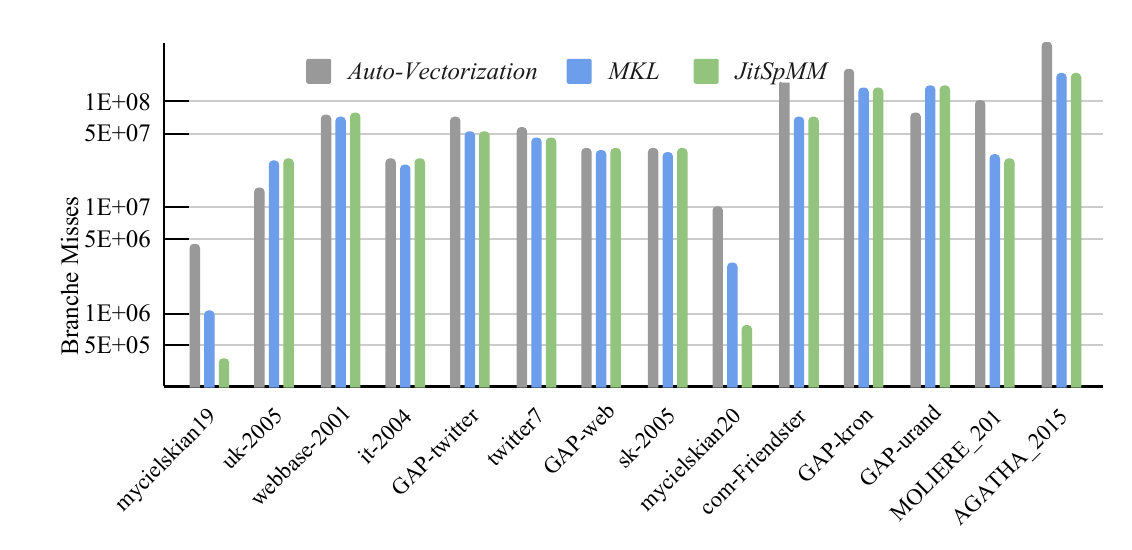}}
    \subfigure[Number of executed instructions.]{\includegraphics[width=0.49\linewidth]{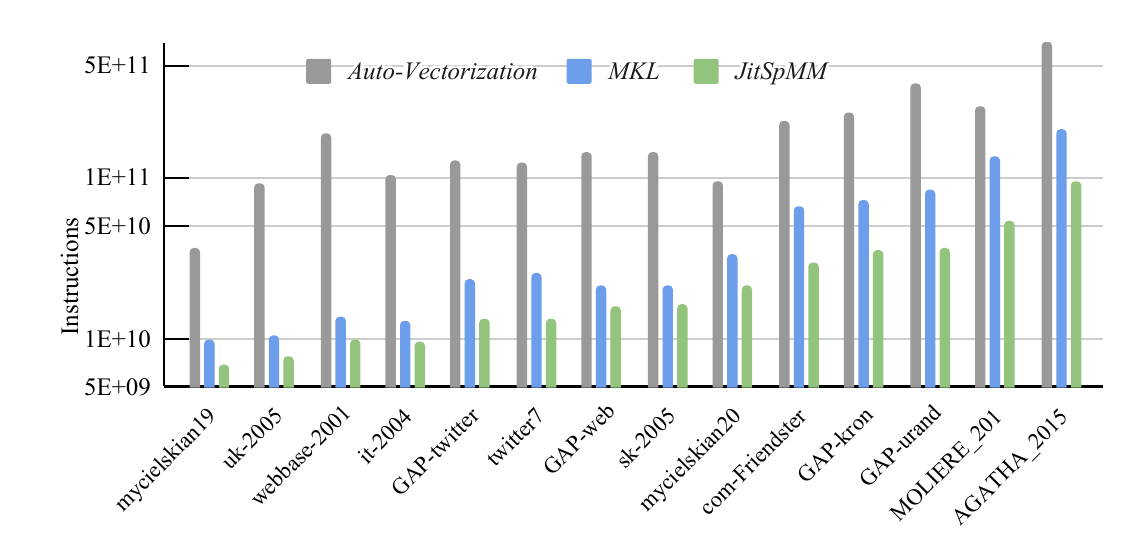}}
    \caption{The profiling results (in log scale) of auto-vectorization, Intel MKL and \name with column number of 16 for input dense matrix. Lower bars represent better performance.}
    \label{fig:profile}
\end{figure*}

To gain a deeper insight into the comparison between our {\name} framework and the baseline AOT solutions, we collected four profiling metrics, which were previously discussed in \cref{sec:jitVSAOTSingleThread}: memory loads, branches, branch misses, and instructions.
We gathered these metrics for all datasets with a column number of 16, as depicted in \cref{fig:profile}. 
In the case of auto-vectorization and {\name}, the profiling outcomes for the three workload assignment methods exhibit high similarity, thereby we use their average values for analysis.
Since Intel MKL's SpMM routine offers only a single implementation, we directly utilize the profiling results of this implementation.
\thomas{What about MKL?}\qiang{MKL only has one implementation.}

First, in terms of memory loads, {\name} consistently demonstrates a large reduction in the number of memory loads across all datasets when compared to the AOT methods.
On average, {\name} employs 2.8\texttimes~and 2\texttimes~fewer memory loads than auto-vectorization and MKL, respectively.
\thomas{Removed the raw values in parenthesis; I think it may not be clear that they refer to the number of memory loads. I think the reduction numbers get the point across.}\qiang{Thanks.}
These results underscore the effectiveness of our register allocation strategy in curbing memory loads through optimized data retention within SIMD registers.
Second, by utilizing runtime information and unrolling the for-loop on line 2 in Algorithm \ref{alg:seqcsr}, {\name} reduces branch instructions by 3.8\texttimes~and 2.9\texttimes, on average, when compared to auto-vectorization and MKL, respectively.
By leveraging runtime information (i.e., the value of $d$ in Algorithm \ref{alg:seqcsr}), {\name} can effectively unroll the loop to mitigate the impact of branch instructions, contributing to improved execution efficiency.
However, {\name} does not exhibit a substantial improvement in reducing branch misses compared to the AOT methods. 
On average, {\name} exhibits 1.4\texttimes~fewer branch misses than auto-vectorization and is on par with MKL.
This phenomenon could be attributed to the high accuracy of the branch predictor within the processor, which tends to forecast correct branch outcomes for the additional branch instructions introduced.
Last, by substantially decreasing unnecessary memory accesses and branch control, {\name} accomplishes the same computation with 7.9\texttimes~and 2.0\texttimes~fewer executed instructions compared to the other two baselines on average.
\thomas{What do you mean by ``equivalent results''? Didn't the framework out perform the other two methods (Figures 9 and 10)? If so, then this should say something like ``Last, {\name} executes 7.9\texttimes~and 2\texttimes~fewer instructions than auto-vectorization and MKL, respectively.
This is possible by substantially reducing the amount of unnecessary memory accesses and branch instructions, which leads to the runtime performance improvements observed in Figures 9 and 10.''}\qiang{I mean {\name} do the same same computation with fewer instruction executed.}
This highlights the efficacy of our approach in optimizing instruction execution by minimizing extraneous instructions associated with memory access and branching.

In summary, by employing a combination of techniques, {\name} successfully addresses the limitations inherent in traditional AOT approaches for SpMM, and the profiling results presented in this section provide a deeper understanding of how {\name} addresses these limitations.
\thomas{I shortened up these paragraphs a bit, so may be worth seeing if we should combine some together if any one of them is too short to be its own paragraph (will be easier to see once the comments are removed).}\qiang{Done.}


\section{Related Works}
\label{sec:rel}
In this section, we provide a brief review of the existing works on the optimization of SpMM, which has been extensively studied over decades on various architectures, including multi-core CPUs and general-purpose GPUs\cite{liu2013efficient,merrill2016merge,rahman2021fusedmm,huang2020ge,yang2018design,hong2019adaptive,hong2018efficient}.
The primary goal of these studies is to enhance SpMM performance by addressing key aspects such as workload balance\cite{merrill2016merge}, optimization of memory accesses\cite{yang2018design,huang2020ge}, and innovative data representation techniques\cite{hong2019adaptive, hong2018efficient}.
The \emph{merge-based} workload division is proposed for SpMV by Merill and Garland\cite{merrill2016merge} to address the workload imbalance of the \emph{row-split}\cite{greathouse2014efficient,ashari2014fast} and \emph{nnz-split}\cite{dalton2015optimizing} methods, using 2-D binary search to find a balanced decomposition of workload.
Yang et al.\cite{yang2018design} generalizes the \emph{row-split} and \emph{merge-split} methods to SpMM on GPUs, and enable coalesced memory accesses for better performance.
Ge-spmm\cite{huang2020ge} is a GPU kernel design for performing SpMM-like operations on sparse matrices represented in the common Compressed Sparse Row (CSR) format, with techniques to ensure efficient coalesced accesses to the GPU global memory and to reduce redundant data loading among GPU warps.
Based on an in-depth analysis to contrast SpMV and SpMM, Hong et al.\cite{hong2018efficient} developed a new sparse matrix representation and computation approach suited to achieving high data-movement efficiency and effective GPU parallelization of SpMM.
Hardware vendor libraries i.e., Intel MKL\cite{wang2014intel} and NVIDIA cuSPARSE\cite{naumov2010cusparse} also provide high-performance (not open-source) SpMM implementation for CPU and GPU.

While these prior studies have made significant strides in accelerating SpMM, it is important to note that they all adhere to the ahead-of-time (AOT) compilation approach. 
Consequently, they encounter, to varying degrees, the limitations that are exemplified in this work.
\thomas{Any prior works on using JIT for sparse matrix operations? Or any kind of assembly code generation/optimization for sparse? Or anything related to SIMD-ifying sparse operations?}\qiang{I didn't find any work using JIT for sparse matrix operations.}

\section{Conclusion}
\label{sec:con}
This work introduced the {\name} framework, which leverages just-in-time (JIT) assembly code generation to enhance the efficiency of Sparse Matrix-Matrix Multiplication (SpMM) operations on modern multi-core CPUs with SIMD extensions. 
By addressing the limitations of traditional ahead-of-time (AOT) compilation approaches, {\name} substantially reduces unnecessary memory accesses, branch operations, and instructions executed.
Our experimental evaluations demonstrate the effectiveness of {\name} across various datasets, showcasing improvements as large as 10\texttimes~compared to baseline AOT solutions. 
\thomas{Anything you can add for future work ideas?}

\bibliographystyle{IEEEtran}
\bibliography{IEEEabrv,references}

\end{document}